\documentclass[11pt]{article}

\usepackage{amsmath}

\newcounter{CommentCtr}

\def\listofcomments{
        \section*{Comments/Corrections\@mkboth{COMMENTS/CORRECTIONS}{COMMENTS/CRRECTIONS}\\{\normalsize Comment number(page number): comment}}
        \begin{description}
        \@starttoc{com}
        \item[\mbox{ }] \mbox{ }
        \end{description}
        \if@restonecol
            \twocolumn
        \fi
}

\newcommand{\PreserveBackslash}[1]{\let\ttemp=\\#1\let\\=\ttemp}

\newcommand{\defeq}{\stackrel{\rm def}{=}}

{\protect\end{itemize}}

{\protect\end{enumerate}}

\newcommand{\TEXfigure}[2]{     
        \begin{center}
        \leavevmode
        \input{tex_figs/#1.tex}
        \centerline{\raise 1em\box\graph}
        \caption{\protect #2}
        \label{#1}
        \end{center}
        \end{figure}
}

\newcommand{\qfalse}{{{\rm \bf f}}}

{\begin{tabular}{lll}}%
{\end{tabular}}
{\begin{supertabular}{lll}}%
{\end{supertabular}}

\usepackage{times}
\usepackage{witsa4}
\usepackage{javaprogram}

\title{A Proposal for Dynamic Access Lists for TCP/IP Packet Filering}
\author{Scott Hazelhurst\\Programme for Highly Dependable Systems\\
School of Computer Science\\
University of the Witwatersrand, Johannesburg\\
Private Bag 3, 2050 Wits, South Africa\\
scott@cs.wits.ac.za}

\date{April 2001}

\newcommand{\denypriority}[1]{\ensuremath{\widehat{\phi_{#1}}}}
\newcommand{\exception}[1]{\ensuremath{{\epsilon_{#1}}}}

\begin{document}
    
\maketitle

\footnotetext{Technical Report TR-Wits-CS-2001-2}

\begin{abstract}\noindent
  The use of IP filtering as a means of improving system security is
  well established. Although there are limitations at what can be
  achieved doing relatively low-level filtering, IP level filtering
  has proved to be efficient and effective.
  
  In the design of a security policy there is always a trade-off
  between usability and security.  Restricting access means that
  legitimate use of the network is prevented; allowing access means
  illegitimate use may be allowed. Static access list make finding a
  balance particularly stark --- we pay the price of decreased
  security 100\% of the time even if the benefit of increased
  usability is only gained 1\% of the time.
  
  Dynamic access lists would allow the rules to change for short
  periods of time, and to allow local changes by non-experts.  The
  network administrator can set basic security guide-lines which allow
  certain basic services only. All other services are restricted, but
  users are able to request temporary exceptions in order to allow
  additional access to the network. These exceptions are granted
  depending on the privileges of the user.
  
  This paper covers the following topics: (1) basic introduction to
  TCP/IP filtering; (2) semantics for dynamic access lists and; (3) a
  proposed protocol for allowing dynamic access; and (4) a method for
  representing access lists so that dynamic update and look-up can be
  done efficiently.
\end{abstract}

\section{Introduction}

The use of IP filtering as a means of improving system security is 
well established. Although there are limitations at what can be achieved 
doing relatively low-level filtering, IP level filtering has proved to 
be efficient and effective~\cite{schuba97}.

The access lists that are used to implement IP filtering contain 
rules that specify which packets should be allowed to pass through 
the firewall. Access lists may last for several years and so may be 
changed from time to time (rules may be added or deleted, old rules 
changed, or the order of the rules change). Nevertheless, an access 
list is relatively static.

The problem with a static access list is that the level of security 
is relatively static. This becomes increasingly a problem as the range 
and type of network travel increases. 

Striking the right balance between usability and security is one of
the key issues in network design. Using static access lists makes
choices in finding a balance particularly stark. Restricting access
means that legitimate use of the network is prevented; allowing access
means illegitimate use may be allowed. A user may only need certain
accesses for 15 minutes a day (1\% of the time), but when they need
the access they really need the access. On the other hand, keeping
access available 99\% of the time when no benefit accrues seems too
liberal. One should only take a risk when some benefit may result. As
an analogy, after I do a large grocery shopping I might leave my car
door and front door wide open while I trudge back and forth carrying
grocery packets because it makes the job easier and faster, but I
certainly don't leave the doors open all the time.

The idea behind dynamic access lists is to allow the rules to change 
for short periods of time, and to allow local changes by non-experts.
The network administrator can set basic security guide-lines which 
allow certain basic services only. All other services  
are restricted. However, users are able to request temporary 
exceptions in order to allow additional access to the network. These 
exceptions are granted, depending on the privileges of the user.
Dynamic access lists have been used in CISCO routers for some
time~\cite{cisco96}. What is being proposed here though is a much more
general framework for making access lists dynamic.

\subsection*{Structure of the paper}

Section~\ref{back:section} gives a basic introduction to TCP/IP 
filtering and explains some of the relevant issues and techniques.
Section~\ref{semantics:section} surveys possible semantics for dynamic 
access lists and proposes one which is argued makes intuitive sense 
and is sound. Section~\ref{protocol:section} presents the outline of 
a proposed protocol for allowing dynamic access. 
Section~\ref{impl:section} describes a method for representing access 
lists so that dynamic update and look-up can be done efficiently.
Section~\ref{exper:section} proposes some experiments to be performed.


\section{Background}

\label{back:section}

\subsection{Firewalling}

Security can be provided at a number of different levels and in 
different places. For example, we may secure individual computers or 
we may secure networks. There are different advantages and 
disadvantages of these different approaches -- see \cite{schuba97} 
for a discussion.  I argue that the advantage of dynamic access lists 
is that it allows more flexibility, allowing defence in depth. In 
addition, we are able to take into account different needs of 
different user classes, rather than just physical location.

Firewalling can be done at different levels. For example, proxies 
use application-layer information in controlling network connections. 
Because they can use high-level information, they are able to make 
good quality decisions. However, this imposes extra costs. 

IP-level filtering is much simpler and therefore cheaper, although 
this limits the intelligence of the filtering.  The use of 
user-classes in the dynamic approach may increase the intelligence of 
the approach.  

Even though the IP filtering is relatively efficient, 
the cost of filtering may still be a significant bottle-neck 
\cite{ballew97}.  Significant work has gone into improving the 
performance of IP filtering 
\cite{gupta99,hazel2000,mchenry97,srinivasan99}. The fact that 
filtering is a bottle-neck means that dynamic filtering must not 
introduce significant extra costs.

\subsection{IP filtering and Rule sets}

TCP/IP filtering is a slightly misleading terminology since in fact it
means filtering using information found in the internet, network
and transport layer headers (depending on the protocol suite).
Typically the information that can be found in the these headers is:
\begin{itemize}
\item source and target addresses of the packet;
\item the protocol of the packet (e.g. udp, tcp, icmp, \ldots);
\item ports;
\item certain flags (for example, a tcp packet contains flags
  indicating status for connection control).
\end{itemize}

\noindent
See a standard reference for more details (e.g.~\cite{washburn96}).

This paper considers TCP/IP packets in particular, but the methods
generalise to similar protocols. 

Filter rules come in several formats; typically these are
proprietary formats. While the expressiveness and syntax of the
formats differ, the following generic description gives a good feeling
for what such rules sets look like. A rule set consists of a
list of rules of the form \(\If{|condition|}{\ |then|\ |action|}\), where
the action is either accept or reject.

\paragraph*{Example: }
A rule in a rule list for a Cisco router~\cite{cisco97} might say
something like:

{\footnotesize
\begin{verbatim}
access-list 101 permit tcp 20.9.17.8  0.0.0.0 
                          121.11.127.20 0.0.0.0 
                          range 23 27
\end{verbatim}
}

\noindent
This says that any TCP protocol packet coming from IP address
20.9.17.8 destined for IP address 121.11.{\linebreak[3]}127.20 is to
be accepted provided the destination port address is in the range
$23\ldots 27$.\qed

\paragraph*{Masking: }
A rule can specify a range of addresses by using \emph{masking} for
both the source and destination addresses (in the above, the masks
were 0.0.0.0, which means no masking). An address is actually a 32 bit
number, which is convenient for humans to express in the quad notation
(four numbers each in the range 0\ldots 255). A mask is expressed
similarly. If a ``1'' appears in the mask, then the value of the
corresponding bit in the address is ignored in matching.
In the above example, since the masks are all 0 the addresses must
match exactly. But, if we had 
\verb/20.9.17.8  0.0.0.255/ as the source address, then any address
with 20.9.17 as a prefix would match. If \verb/20.9.17.8  0.0.0.1/ 
were the source, then any
\emph{even} address with 20.9.17 as a prefix would match.

\paragraph*{Matching a list of rules: }
The rules are searched one by one to see whether the condition matches
the incoming packet: if it does, the packet is accepted or rejected
depending on the action (which will either be accept or reject); if
the condition does not match the rule, the search continues with the
following rules.  If none of the rules matches, the packet is rejected.

Since the rules are checked in order, the order in which they are
specified is critical. Changing the order of the rules could result in
some packets that were previously rejected being accepted (and/or
\emph{vice-versa}). 

This paper uses CISCO access-list format as the basis specifying the
rule set, but the methods proposed generalise to other formats.


\section{The Semantics of Exceptions}

\label{semantics:section}

The complication in defining the semantics arises from the importance of 
the order of an access list to the semantics of the list. This section 
explains why this is a problem, presents alternate semantics, proposing 
one of them as possible semantics of exceptions.

\subsection{The naive semantics}

\label{semantics:section:naive}
The naive semantics would be to consider the exceptions as an extension of 
the base list. In this view, the exceptions are temporarily pre-pended 
or appended to the base list. While this is simple to implement and in some 
cases could have an understandable effect, it is highly problematic.

Suppose the extensions are pre-pended to the base list. Then, because 
ordering is critical, all extensions will over-ride all base rules: there 
is no way of ensuring that certain rules are always obeyed. 

Conversely, if the rules are appended then an exception can never
over-ride any of the base rules: in effect, an exception is only an
exception to the default reject rule.  This is more desirable and useful
than prepending, but it reduces the flexibility of a security policy, and
would be likely to lead to a situation where the base rules are more
liberal than necessary in order to allow the exceptions to have some use.

More sophisticated direct modification of the access lists seems 
unlikely given the high-level of expertise required to make changes 
and the complexity of the interplay between rules. Inserting dynamic rules 
somewhere in the middle of the list may give more flexibility but will 
not be flexible enough, and be too difficult to understand the effect.

\subsection{A tree-based approach}

\label{semantics:section:tree}

A more sophisticated policy is to have multiple lists, instead of
having one access list.  These lists could be ordered by priority, and
for each list, rejects could be mandatory or flexible.  The highest
priority list or the base list would be the main list which the
network administrator sets up.  In practice, the lower-priority lists
would be of a fairly simple structure: a number of accepts followed by
some mandatory and flexible rejects.

In the tree-based approach, in order to decide 
whether a packet should be filtered, the lists are searched in order.
If, while searching a list, an accept rule is matched, the packet is 
accepted; if a mandatory reject rule is matched the packet is 
rejected; otherwise if a flexible reject is matched the next list of 
rules in priority  is searched.

Dynamic access requests can be effected by pre-pending the access 
rule to one of the lower-level priority lists. Different security 
policies can be implemented by allowing different classes of user to 
pre-pend to different lists (the higher the priority of the list 
to which the user can prepend, the more privileged the user), and by 
the choice of the reject rules in the lower-priority lists. 

This policy has intuitive appeal, and could be implemented relatively 
efficiently.  However, it has a very serious
problem in that it alters the semantics of rule lists significantly;
in particular, in the interplay between the ordering of the rules
within a list and the ordering of lists. The simple example lists 
shown in Figure~\ref{semantics:fig:nonmono} illustrates the problem. 

In  Figure~\ref{semantics:fig:nonmono}, we have two access lists,
with List 0 being of higher priority 
than List 1. Condition $c_{i,j}$ gives the matching condition for a 
packet for rule $j$ in List $i$. Suppose we wish to know whether to 
accept or reject packet $p$ and assume that packet $p$ matches 
conditions $c_{0,1}$, $c_{0,2}$ and $c_{1,0}$ but does not match the
other conditions. By the semantics just 
explained, $p$ will be accepted; first List 0 is searched and since 
the first rule that matches $p$ is rule 1, a flexible reject, we carry 
on searching List 2, and since the first rule that matches in List 1 
is rule 0, we accept.

However, suppose that we delete rule 1 of List 0. Now, the packet $p$ 
would be rejected since the first rule that $p$ matches against will 
be rule 2, a mandatory delete. Thus, we get a situation where 
deleting a reject rule can increase the probability of a packet being 
rejected\footnote{And, thinking of the flexible reject as a
  conditional accept doesn't help either -- opposite monotonicity
  constraint violations can be shown.}.

\begin{figure}[ht]
\begin{tabular}{|l|l|l|}\hline
   Rule \# &List 0 & List 1  \\\hline
   0 & accept $c_{0,0}$           & accept $c_{1,0}$ \\
   1 & reject flexible  $c_{0,1}$ & reject flexible $c_{1,1}$ \\
   2 & reject mandatory $c_{0,2}$ & reject flexible $c_{1,2}$ \\\hline
\end{tabular}
\caption{Sample lists}
\label{semantics:fig:nonmono}   
\end{figure}

The underlying problem that this example illustrates is that the order 
of rules in a list is very important for the semantics. Implementing 
this `tree' semantics confuses issues because ordering is used both 
to establish priority between lists, and to implement the standard 
semantics of rule lists. It makes backwards compatibility with 
existing rule lists difficult.

\subsection{Principles of semantics}

\label{semantics:section:principles}

The issues raised sections \ref{semantics:section:naive} and 
\ref{semantics:section:tree} lead to the following principles relating 
to the semantics of dynamic lists:

\begin{enumerate}
    
    \item The semantics should support flexible policies.
    
    \item The behaviour of dynamic changes and the interaction between 
    the dynamic rules and the base rules must be clearly 
    understandable to the person maintaining the base list, without 
    knowledge of what the dynamic rules are.
 
    \item It must be possible for a user to request a dynamic access 
    rule without knowledge of the base rules or other dynamic rules.
    
    Of course, the firewall may not allow such a request if it 
    conflicts with security policy, but a non-expert user (or a simple 
    agent on behalf of a  user) should be able to make  the requests.
    
    \item The base and dynamic rules must be well-behaved. Adding an
    \emph{accept} rule should increase the chance of packets being 
    accepted; adding a \emph{reject} rule (flexible or mandatory) 
    should increase the chance of packets being rejected. (And 
    conversely for deleting rules).
    
    \item Ideally, we should be able to use existing rule sets 
    without much change.
\end{enumerate}

\noindent
Efficiency is obviously also an important question, addressed in 
Section~\ref{impl:section}.

\subsection{The multiple list priority-based approach}

\label{semantics:section:multiplepriority}

The multiple list priority-based approach also uses multiple rule
lists, ordered by priority.  Again, for each list the reject rules are
split into mandatory or flexible.

The security policy is succinctly expressed as:

\begin{quote}
    A packet $p$ is accepted by the firewall if it is accepted by 
    rule list $j$ and there exists no mandatory reject rule in 
    lists $0\ldots j-1$ that match it.
\end{quote}

\noindent
Again, dynamic changes can be effected by allowing users to prepend 
accept rules to one of the lower-priority lists, depending on the users' 
privilege level.

These semantics are far clearer than the semantics in 
Section \ref{semantics:section:tree} and more closely meet the conditions
of Section \ref{semantics:section:principles}.

Implementation of this approach is not as simple as that of Section~ 
\ref{semantics:section:tree}, but the implementation technique 
described in Section~\ref{impl:section} would allow an efficient 
implementation.

The disadvantage of this approach is that it would require the 
duplication of rules. For example suppose that we wish to have as a 
default or base rule that access to machine $x$ is not allowed, but 
that users in class 1 to $i$ should be allowed to over-ride this rule, 
and users in class $i+1$, \ldots should not.  Implementing this 
requires flexible reject rules in  the base list and in lists 
$1\ldots i-1$, and a mandatory reject rule in list $i$. This 
proliferation of rules will make administration much more difficult 
and increase  the chance of errors.

\subsection{The group-based approach}

\label{semantics:section:group}

This paper proposes the group-based approach as the 
semantic basis for supporting dynamic access lists. 

In the group-based approach (GBA), there is one base access list and
$n$ exception lists (numbered $0\ldots n-1$, assuming $n$ user
groups). The base access list is the existing access list, except that
each reject rule has a set of associated group identifiers $0\ldots
n-1$. These group identifiers indicate which groups are able over-ride
the reject rule.  Groups may be contained within other groups.  The
semantics of the GBA are:

\begin{quote}\label{semantics:quote:gba}
    A packet $p$ is accepted by the firewall if
    \begin{enumerate}
         \item It is accepted by the base list; or
         \item It is accepted by some exception list $j$ where all 
         reject rules in the base list are labelled by $j$ or by a 
         super-group of $j$.
     \end{enumerate}
\end{quote}

\noindent
Thus, if a rule in the base list is not labelled with any group 
identifier then it cannot be over-ridden.

I argue that this semantics meets the conditions of 
Section~\ref{semantics:section:principles}.

\begin{enumerate}
       \item Flexible policies are supported since by assigning
       several priority levels for reject rules, the network
       administrator can allow different classes of user different
       abilities to request dynamic access.
    
    \item Interaction between the base list and the exceptions is 
    clear since (a) the normal semantics of the access lists is 
    disturbed very little; and (b) the use of group membership for the 
    reject rules makes it clear what different classes of users can 
    and can't do.
 
    \item Users do not need to know what the base rules are in making 
    a request for dynamic access.
    
    \item The monotonicity constraints are met.
    
    \item Existing rule sets can be used as-is.  Dynamic access can
    then be implemented over time by assigning group membership. 
    There is no need for radical change.
\end{enumerate}

\noindent
The efficient implementation of the GBA is discussed in 
Section~\ref{impl:section}.

\subsubsection{Example}

\label{semantics:section:example}

Here we look at a simple example shown in
Figure~\ref{semantics:section:dynex1}.  The format used is similar to
the CISCO access list format; the number at the beginning of each line
is just used for reference in this explanation.  The example gives
rules for a subnet 128.128.128.  The machine 128.128.128.15 is a
special server; the other machines in the range 128.128.128.0 to
128.128.128.127 are for staff; machines in the range 128.128.128.128
to 128.128.128.255 are for students.  In this example, group 0 is the
\emph{staff} group, group 1 the \emph{student} group, and group 2 the
\emph{all} group, which contains both groups 0 and 1.

\begin{figure}[ht]
{\footnotesize
\begin{verbatim}
1: accept tcp 0.0.0.0 255.255.255.255   128.128.128.15 0.0.0.0  eq 88
2: deny tcp 0.0.0.0 255.255.255.255   128.128.128.15 0.0.0.0   
3: accept tcp 0.0.0.0 255.255.255.255   128.128.128.0 0.0.0.255 eq 88
4: accept tcp 0.0.0.0 255.255.255.255   128.128.128.0 0.0.0.255 ge 32000 
5: deny  tcp 0.0.0.0 255.255.255.255   128.128.128.0 0.0.0.255 range 0 87
6: deny 0 tcp 0.0.0.0 255.255.255.255   128.128.128.0 0.0.0.127 ge 89
7: deny  tcp 0.0.0.0 255.255.255.255   128.128.128.128 0.0.0.127 lt 16000
8: deny  1 tcp 0.0.0.0 255.255.255.255   128.128.128.128 0.0.0.127 ge 16000
9: deny 2 everything
\end{verbatim}
}
\caption{Simple example dynamic base list}
\label{semantics:section:dynex1}.
\end{figure}

Rule 0 says that we accept tcp connections to the machine 15 on port
88.  Rule 1 says that we deny (with no exceptions allowed) all tcp
connections to machine 15 on any port.  Since rules are examined in
order the combined effect of these two rules is that tcp connections to
machine 15 are accepted on port 88 only.  Rules 3 and 4 say that on
all machines on the subnet we accept tcp connections on port 88 and
ports in the range 32000 to 65535.  (Since rules 1 and 2 come before
rules 3 and 4, rules 3 and 4 will not affect machine 15 since we have
denied access to ports other than port 88 on machine 15) Rule 5 denies
tcp connections to any machine on the subnet to any
port in the range 0 through 87 and no derogation from this rule is allowed. 

Rule 6 says that we deny tcp connections to any of the staff machines
(range 128.128.128.0 to 128.128.128.127) on any port numbered 89 or
higher.  However, we allow members of the staff group to request
exceptions to this rule.  Note there is some overlap between this rule
and previous rules: in this overlap the previous rules take priority
because they come first.

Rule 7 says that we deny tcp connections to any student machines on
any port numbered less than 16000.  (Since rule 3 comes first,
connections to port 88 are still allowed).  Rule 8 says that we deny
any connections to student machines on ports with a number greater
than or equal to 16000.  Members of the student group are allowed to
request exceptions to this rule. (Connections to ports $\geq 32000$ are
still allowed since rule 4 comes first.)

Finally, rule 9 denies any other packets. However, any member of 
either the staff or student group can ask for exceptions.

Note, that without any exceptions, the semantics of the access list 
is unchanged from the standard semantics.

Now, let's look at the effect of exceptions.
Suppose we have the following exception lists:

{\footnotesize
\begin{verbatim}
   Exception list 0
   0.0 accept tcp 0.0.0.0 255.255.255.255  128.128.128.1  0.0.0.0  eq 100
   0.1 accept tcp 0.0.0.0 255.255.255.255  128.128.128.1  0.0.0.0  range 0 90
   0.2 accept tcp 0.0.0.0 255.255.255.255  128.128.128.129  0.0.0.0  eq 16000

   
   Exception list 1
   1.0 accept tcp 0.0.0.0 255.255.255.255  128.128.128.2  0.0.0.0  eq 100
   1.1 accept tcp 0.0.0.0 255.255.255.255  128.128.128.129  0.0.0.0  eq 16000
   1.2 accept icmp 0.0.0.0 255.255.255.255  128.128.128.129  0.0.0.0  

   Exception list 2
   2.0 accept tcp 0.0.0.0 255.255.255.255  128.128.128.130  0.0.0.0  eq 16000
   2.1 accept icmp 0.0.0.0 255.255.255.255  128.128.128.130 0.0.0.0  
\end{verbatim}
}

Exception 0.0 allows tcp connections to be made to port 100 on machine 1.
Any packets that come to this port will be accepted because they are 
accepted by exception list 0 and the only rules that deny access to 
this port are rules 6 and 9. Both allow exceptions to be requested by 
members of group 0 and so this exception can be honoured. 

Exception 0.1 purports to allow tcp exceptions to ports 0 through 90
on machine 1.  However, packets that come to ports 0 through 87 will
not be enabled by this exception since they are rejected by rule 5
which does not allow derogations.  TCP packets going to port 88 were
in any event allowed by rule 3.  Packets going to port 89 and 90 will
be allowed by the exception since the rules that deny access to this
port (6 and 9) allow exceptions to be made by members of group 0.

Exception 0.2 purports to allow packets going to port 16000 on 
machine 129. However, rule 8, which applies here, can only be 
derogated from by members of the student group.

Exception 1.0 purports to allow  tcp connections to port 100 on 
machine 2. This has no effect, however, since rule 6 denies access to 
this port and as the rule only allows derogations by staff member, 
only exceptions in exception  list 0 can over-ride it.

Exception 1.1 does have effect though since it allows access to port
16000 on machine 129 and the relevant deny rule (8) has is labelled
group 1 (and hence can be over-ridden in exception list 1). 
Similarly, exception 1.2 will allow icmp packets to reach machine 129
since the relevant deny rule (9) is labelled 2 (the \emph{all} group).

Exception 2.0 purports to allow tcp connections to port 16000 on 
machine 130. However, as rule 8 denies such connections and is 
labelled group 1, exceptions that over-ride rule 8 must be in exception 
list 1. However, exception 2.1 does allow icmp access to machine 
130 since rule 9 is labelled 2.

\subsubsection*{A priority based approach}

Note that if the groups are set up in a strictly hierarchical way, 
the group-based approach corresponds to a priority-based one.

\subsection{Should \emph{accept} rules have priorities?}

\label{semantics:section:accept}

Should accepts also have priorities and allow exceptional denies? This 
has the appeal of symmetry, but while this would allow extra 
functionality it is not clear that this would be useful and would 
certainly lead to undesirable problems of interactions between users.


\section{Protocol for dynamic update}

\label{protocol:section}

This section proposes a protocol for dynamic access list by 
specifying how communication between user processes and the firewall 
takes place. At this stage this is only a preliminary proposal since 
we need to get experience with more substantial prototype systems 
before finalising details. When a user\footnote{Here, \emph{user} 
will probably be some process acting on behalf of a human, rather than a 
human.} wishes to ask for dynamic access, it sends a request to the 
firewall. The firewall logs and validates the request, checks to see 
whether the request can be fulfilled and then responds to the user.
This section describes this communication in more detail. How the 
firewall maintains the requests, performs updates, and does look-ups 
is covered in Section~\ref{impl:section}.

\subsection{Request for dynamic access}

A request for dynamic access takes four steps:
\begin{itemize}
    \item The user sends a request to the firewall asking for access;
    \item The firewall sends back a response indicating to what 
    degree the access can be given;
    \item The user then sends a message to confirm whether it wants 
    the access given.
    \item If the firewall receives the confirmation within a given 
    time interval, the exception is made; otherwise the exception is 
    not made.
\end{itemize}

\paragraph*{First step -- asking for access: } The user sends a packet 
to the firewall asking for access to be given. The packet is sent as a 
UDP packet to a well-known port on the firewall, and contains the 
following information:
\begin{itemize}
    \item Originating IP address and UDP port (part of the IP/UDP 
    headers).
    \item User identification: this must enable the firewall to 
    identify the  group of the user.
    
    \item Access required: this would be a list of accept rules 
    indicating what access is wanted.
    
    \item Expiry time: The user specifies for how long the exception 
    should be enabled.
\end{itemize}

\paragraph*{Firewall response: } When the firewall receives the 
\emph{request} packet, it validates the user and determines the 
group membership. It then examines the update request.
\begin{itemize}
    \item If the update request can be met completely (i.e. all the
    requests asked for do not clash with a deny rule which cannot be 
    derogated from by members of the user's group), then the
    firewall sends back to the user an \emph{allow full} message with
    a unique dynamic update ID.
    
    At the same time, the firewall inserts the update in a queue of
    pending exceptions.
    
    \item If the update request cannot be met at all (i.e. all the 
    updates clash with a deny rule which cannot be derogated from by
    the user under  GBA) a \emph{reject} 
    message is sent back to the user.
    
    \item In general, the update firewall may be able to honour some 
    of the dynamic update requests but not others (see the discussion 
    on Exception 1.0 in Section~\ref{semantics:section:example}). In 
    this case the firewall sends back to the user an \emph{allow partial} 
    message with a unique dynamic update ID, as well as a description 
    of those requests that can be honoured.
    
    At the same time, the firewall inserts the update in a queue of
    pending exceptions.

\end{itemize}

\noindent There are number of open implementation decisions: should
the firewall accept update changes from outside or only from inside? 
How are users identified?  It may be the userid of a user that the
firewall knows about (it could have its own database or use NIS), or
it could be a capability-based system.  The identification could be
authenticated by digital signature. This and other messages in the 
exchange may or may not be encrypted.

What information is returned with the \emph{allow partial} message is 
another implementation decision. In the approach proposed in 
Section~\ref{impl:section} a full description will be returned. 
However, there might be reason just responding with a 
\emph{reject} or \emph{allow}.

\paragraph*{Confirmation by user: } If the user receives a 
\emph{reject} message, then it has to reconsider what it wants.
If the user receives an \emph{allow} message, it decides whether it 
wishes to use the update, and if so it sends back to the firewall a 
\emph{confirm} message together with the unique ID. This step ensures 
that the firewall only implements changes that the user really wants. 
It also helps reduce the chance of spoofing attacks on the firewall.
    
\paragraph*{Firewall implements the exception: } When the firewall 
receives the \emph{confirm} request it removes the update request from 
the pending queue, and adds it to the appropriate exception list, 
timestamping the update as it does so.

Periodically, the  update queue is scanned and old requests are purged.

\subsection{Undoing a an update}

Since the access list is supposed to be dynamic, it must be possible 
to undo the change. This can be done easily by deleting the 
exception. There are two proposed mechanisms for this:

\begin{itemize}
    \item The user sends a \emph{delete} request to the firewall with 
    the unique ID of the update. The firewall authenticates the 
    request and then deletes the update.
    
    \item The firewall periodically checks the exceptions and when an 
    exception has expired (the time since it was added to the 
    exception list longer than the expiry time associated with the 
    exception), the exception is deleted.
\end{itemize}

\noindent
Other mechanisms are possible. For example, the firewall could 
monitor the traffic associated with the exception and when it detects 
that there has been no traffic for some period then the exception is 
deleted. However, this is likely to be considerably more complex and 
heavy-weight than the proposed method here.

\subsection{Renewing an exception: } A  user may wish to extend the 
life-time of an existing exception. This can be done by sending the 
\emph{renew} message to the firewall with user ID and expiry time.


\section{An implementation mechanism for dynamic lists}

\label{impl:section}

The following principles guide the implementation mechanism proposed 
here.  The principles are listed in descending order of importance:
\begin{itemize}
    \item The cost of doing look-up for packets that are not affected 
    by dynamic update rules should not be adversely affected;
    \item For each update rule, at least 10 additional packets will 
    be affected by the update rule; therefore, the cost of look-up for 
    exceptional packets must be small;
    \item The cost of updating and undoing updates must be small.
\end{itemize}

This section is structured as follows: Section~\ref{impl:section:basic} 
discusses the basic method for representing access lists; 
Section~\ref{impl:section:updates} presents the method of efficiently 
performing updates to the access list and performing lookup; and 
Section~\ref{impl:section:undo} shows how undoing updates can be 
performed.

\subsection{Basic Representation of Access Lists}

\label{impl:section:basic}

The chosen method for representing an access list is as a single
(rather large) boolean expression.  As described in
Section~\ref{back:section} each rule in the access list is a condition
on the bits in the packet header, and hence if we represent each bit
in the packet header with a boolean variable, we can represent the
condition as a boolean expression over these variables.  Given a
packet to filter, we give the variables in the expression their values
as given by the bits in the packet header.  The packet is accepted
exactly when the boolean expression evaluates to true.

\paragraph*{Constructing a boolean expression for each rule: } The
boolean expression for each rule is constructed from its constituent
components.  For example, if the protocol is represented in 8 bits in
the packet header, then we introduce 8 boolean variables
$p_{0},\ldots, p_{7}$.  We can then represent the condition that the
packet must be a packet of protocol type protocol 1 by the boolean
condition:

$$p'_{7}p'_{6}p'_{5}p'_{4}p'_{3}p'_{2}p'_{1}p_{0}$$

\noindent
where juxtaposition is used for conjunction and primes represent
negation. 

\paragraph*{Constructing a boolean expression for the list: }
An entire access list can be thought of as a big if-then-else 
statement, and so once we have converted each individual rule into an 
a boolean expression it is a straight-forward matter to convert the 
entire list into a boolean expression.

For example, suppose we have the list:
{\footnotesize
\begin{verbatim}
1: accept tcp 0.0.0.0 255.255.255.255   128.128.128.15 0.0.0.0  eq 88
2: deny 0 tcp 0.0.0.0 255.255.255.255   128.128.128.15 0.0.0.0   
3: accept tcp 0.0.0.0 255.255.255.255   128.128.128.0 0.0.0.255 eq 88
\end{verbatim}
}

Suppose the boolean expressions associated with rules 1, 2, and 3 
are $\phi_{1}, \phi_{2}$, and $\phi_{3}$ respectively. Then the entire 
list can be represented by the expression:
  $\phi_{1}\lor \lnot \phi_{2}\land \phi_{3}.$
  
The detailed mechanics of this translation are beyond the scope of
this paper.  The important point is that these boolean expressions can
be efficiently represented using binary decision
diagrams~\cite{bryant92}.  See~\cite{hazel98a, hazel2000} for a
detailed explanation of how this is done.  The work of
Attar~\cite{attar2001} and Sinnappan~\cite{sinnappan2001} shows that
this method of representing access lists is a very compact
representation and that look-up can be performed competitively
(with respect to other methods) in both software and hardware.  What
particularly made this representation scheme appear promising for
dynamic access lists were the results that (a) the cost of lookup is
robust in the number of access rules and (b) the variation on look-up
cost is very low.

\subsection{Implementing updates and exceptions}

\label{impl:section:updates}
    
\subsubsection{Representation of base list and exceptions}

The previous section showed that any set of access list conditions 
can be represented as a boolean expression. This section describes how 
the base list and exceptions can be represented under GBA.

The following notation is used:

\begin{itemize}
    \item $\phi_{B}$: The boolean expression representing the base access 
    list. 
    
    If we are not considering exceptions, then a packet is accepted 
    by  the list if under the interpretation of variables given by 
    the bits in the packet $\phi_{B}$ is true.

    \item $\phi_A$: The boolean expression representing the access list
      together with exceptions. Where there are no exceptions,
      $\phi_A=\phi_B$. 

    \item $\denypriority{i}$: 
     for each group $i$, the disjunction of the
        \emph{deny} rules that are not labelled with $i$ or a
        supergroup of $i$.  So, if we instantiate the variables in
        $\denypriority{i}$ with values from the bits in a packet
        header, we find out whether an exception in list $i$ can
        over-ride \emph{reject} rules in the base list in order to
        allow this packet.

    \item $\exception{j}$: the expression representing exception list
    $j$.

\end{itemize}

\noindent Under the GBA semantics given in
Section~\ref{semantics:section:group} (on 
page~\ref{semantics:quote:gba})  the boolean expression that
represents a prioritised access list and $n$ exception lists is:

\begin{equation}
\phi_{A} \defeq \phi_{B}\;\lor\;(\lor_{i=0}^{n-1}\lnot 
\denypriority{i}\land\exception{i})
\label{impl:eqn:phiA}
\end{equation}

\noindent
The firewall keeps the following information:
\begin{itemize}
    \item $\phi_{B}$, the representation of the base list;
    \item For each update request $u$, $\phi_{u}$, the boolean 
    expression representing the exception, the ID of the request, 
    expiry time and the group of the originator of the request; 

    The update requests are stored in a manner so that they can 
    efficiently be accessed by ID number, by expiry time, and by
    group number.
    
    \item For each $i$, $\denypriority{i}$;
    \item For each $i$, $\exception{i}$;
\end{itemize}

\subsubsection{Making an update}

This section describes in detail the behaviour of the firewall in 
response to the protocol described in Section~\ref{protocol:section}.

\paragraph*{Update request received: }
When a new update $u$ arrives with originator priority $j$, the 
following is done:
\begin{itemize}
    \item $\phi_u$ is computed, the boolean expression representing $u$;
    \item $\phi_u \land \lnot\denypriority{j}$ is computed: This represents
        which exceptions that a user of this level of priority can be
        granted. 
      
        \begin{itemize}
        \item If $\phi_u \land \lnot\denypriority{j} = \qfalse$, then
          none of the exceptions can be granted;

        \item If $\phi_u \land \lnot\denypriority{j} = \phi_u$, then
          all the exceptions can be granted;
        \item Otherwise only some of the requests can be granted.
        \end{itemize}
 
    \item If none of the exceptions can be granted, the request is
      deleted and the user is sent a \emph{reject} message.

    \item Otherwise the user is sent an \emph{allow full} or
      \emph{allow partial} message, and the request is put on the
      pending list.

      In the case of an \emph{allow partial} message various types of
      information can be returned to the user. The most useful would
      be tabular representation of $\phi_u \land
      \lnot\denypriority{j}$. An algorithm for presenting this is
      described in ~\cite{hazel2000}.

\end{itemize}

\paragraph{Confirm message received: } When the firewall receives a
      \emph{confirm} message for update $u$, the following happens:
      \begin{itemize}
      \item The update is stored with the associated information in an
      easily accessible form as described above; 
      \item $\exception{j}$ is updated: 
      $\exception{j} \leftarrow
      \;\exception{j}\lor\phi_u\land \lnot\denypriority{j}$; 
      \item $\phi_A$ is recomputed according to
      Equation~\ref{impl:eqn:phiA}.
      \end{itemize}

\subsection{Undoing updates}

\label{impl:section:undo}

In principle, undoing an exception is straight-forward. When an 
exception of priority $j$ is undone, 

\begin{itemize}
    \item The exception is removed from the list of active exceptions;
    \item  $\exception{j}$ is recomputed;
    \item $\phi_A$ is recomputed according to
       Equation~\ref{impl:eqn:phiA}. 
\end{itemize}

Undoing is likely to be more expensive than creating the exceptions 
because there is much more computation to be done. When we make an 
exception, there are relatively few changes to be made, whereas with 
undoing there will be many more.  A number of factors will, hopefully, 
reduce the significance of this:

\begin{itemize}
    
    \item Undoing is unlikely to be time-critical (i.e. if it takes a 
    few seconds extra to undo the exception, there will be no serious 
    consequences). Thus, we only need to consider the cost in terms 
    of the load it places on the server. And since it is not critical 
    undoing can be done as a low-priority process.
    
    \item BDD packages use caching of results to get performance 
    improvement. Undoing may be able to exploit this.
    
\end{itemize}

However, the cost of undoing is likely to be the critical factor in 
the performance of this approach.


\section{Conclusion and Proposed experiments}

\label{exper:section}

This report has proposed the use of dynamic access lists for IP 
filtering, arguing that the benefits of dynamic lists are increased 
flexibility and security.

A semantics for dynamic access lists was proposed and motivated, as 
well as a protocol for interaction between the users and the firewall.

The previous section described a mechanism for implementing dynamic 
access lists.

\subsection*{Proposed experiments}
The key question in evaluating the proposed protocol and
implementation is time performance, and to a lesser extent memory
(though previous work has shown that memory is unlikely to be a
constraint).

In terms of performance there are two criteria by which this must be 
measured:
\begin{itemize}
    
    \item Increased latency experienced by users;
    
    \item Extra work-load placed on the firewall
\end{itemize}

Previous work indicates that dynamic access lists should not lead to 
increased look-up time as the boolean representation is a robust one. 
Hence, from a  user perspective, latency only 
increases when an update request is being made, and not when packets 
are being filtered. Since update requests are relatively rare, and 
given the wide variance of time in internet access, for the purpose 
of this work, a  time of 5s cost from start of request to 
request being granted was set upon. This would be indistinguishable 
from other network delays to the average user.

The firewall may have to deal with requests from many users. If the 
cost of making and undoing updates is significant then the performance 
of the firewall as a filter may degrade. Thus, we are interested in 
the overhead from the firewall's point of view. This then will be the 
central thrust of the experiments.

\paragraph*{Acknowledgements: } Pekka Pihlajasaari made a number of 
valuable comments on a draft of this material, including the 
suggestion of generalising from a priority-based to a group-based 
scheme.

\bibliography{refs}

\end{document}